\documentclass[]{aa}
\usepackage{graphicx}
\usepackage{natbib}
\usepackage{ulem}
\begin{document}

\title{Multiline Zeeman signatures as demonstrated through the 
Pseudo-line} 

 \author{M. Semel\inst{1}, J.C. Ram\'{i}rez V\'elez\inst{1},
   M.J.~Stift\inst{2,3},  M.J. Mart\'{i}nez Gonz\'alez\inst{3},
   A. L\'opez Ariste\inst{4} \& F. Leone\inst{5}}

\institute{
LESIA, Observatoire de Paris Meudon. 92195 Meudon,
France. \email{Meir.Semel@obspm.fr, Julio.Ramirez@obspm.fr} 
\and Institute for Astronomy, Univ. of Vienna, T{\"u}rkenschanzstrasse 17,
1180 Vienna, Austria.
\and LUTH, Observatoire de Paris Meudon. 92195 Meudon, France. \email{martin.stift@univie.ac.at, Marian.Martinez@obspm.fr} 
\and THEMIS, CNRS UPS 853, c/v\'{\i}a L\'actea s/n. 38200. La  Laguna,
Tenerife, Spain; \email{arturo@themis.iac.es}
\and INAF â€" Osservatorio Astrofisico di Catania, Via S. Sofia 
n 78,95123 Catania, Italy
}

\offprints{meir.semel@obspm.fr}

\titlerunning{Multiline Zeeman Signatures }
\authorrunning{Meir Semel et al.\ \ }
 \date {Received date , accepted date}

\begin{abstract}
{In order to get a significant Zeeman signature in the polarised
spectra of a magnetic star, we usually 'add' the contributions of numerous
spectral lines; the ultimate  goal is to recover the spectropolarimetric
prints of the  magnetic field in these line additions.}  
{Here we want to clarify the meaning of these techniques of line
addition; in particular, we try to interpret the meaning of the 
'pseudo-line' formed during this process and to find out why and
how its Zeeman signature is still meaningful.}
{We create a synthetic case of lines addition and apply well tested
standard solar methods routinely used in the research on magnetism 
in our nearest star.}
{The results are convincing and the Zeeman signatures well detected;
Solar methods are found to be quite efficient also for stellar 
observations.}
{The Zeeman signatures are unequivocally detected in this multiline 
approach. 
{\bf We may anticipate  the outcome magnetic fields to be  reliable} 
well beyond  the weak-field approximation. 
Linear polarisation in the spectra of 
solar type stars can be detected when the spectral resolution is 
sufficiently high.}
\end{abstract}


\maketitle

\section{A general introduction to ZDI and PCA}

In the 1980s it became clear that cool stars, mainly rapid rotators,
exhibit solar type activity. Rapid rotation is one of the essential
ingredients to the {\it stellar dynamo}. While there is not yet any
completely satisfactory theory or model of the {\it solar or the stellar
dynamo}, it is quite clear what drives such a process, viz. convection,
rapid rotation and differential rotation. Tentatives to detect the
Zeeman effect in these active stars however failed for 2 main reasons:
\begin{itemize}
\item The magnetic configuration of the stellar magnetic fields can
be quite complex. The simultaneous appearance of opposite magnetic
polarities may lead to cancellation of the respective contributions
to the integrated polarisation signal in the stellar spectrum.
\item The broadening of the spectral lines due to even moderate stellar
rotation rates can become a serious handicap when measuring the
magnetic field of the star.
\end{itemize}
Consideration of these two aspects was at the object of paper~{\sc i}
(Semel 1989) where the term ZDI stood for the detection of a Zeeman
signature thanks to the Doppler effect. The latter can help to
disentangle the contributions from opposite magnetic polarities and
their respective opposite polarisation which otherwise may cancel.
Thus the combination of Zeeman and Doppler effects is one of the
reasons why the detection of the Zeeman effect in fast rotating active
solar type stars becomes possible. Nowadays however the term ZDI is
often understood as magnetic mapping of the stellar surface, based on
the inversion of Stokes profiles (frequently only $I$ and $V$) sampled
over a full period of rotation.

We therefore propose the term {\em Multiline Zeeman Signature}
(henceforth MZS) to denote a method for {\bf just} the detection of a mean
Zeeman signature, using numerous spectral lines.
We shall not attribute any direct physical meaning to the MZS.
We simply require the application of an operator {\sf O} (or a detector 
{\sf O}) to a
polarised spectrum, creating a particular $MZS_O$. The inversion code
then should use exactly the same operator {\sf O} for the comparison and 
inversion of
the observational data in order to get the desired information on the
magnetic field. We leave this demonstration to a forthcoming paper.

  {\bf In this paper we discuss the problem of {\it Zeeman signatures} in the 
polarized spectra of solar type stars. It may concern any of the three 
components of the magnetic vector, at any stellar point and in any of the 
four components of the Stokes 
vector in the spectrum of the particular  stellar point. 
( and the SUN as a special case).

 This paper does not treat {\it inversion}. This last is postponed 
to the forthcoming papers.
By {\it inversions} we mean recovering not only  magnetic field but also the 
model of atmosphere !

This paper does not resolve {\it mixtures of polarities}. It is left for 
later steps.}

We therefore turn back  to the more elementary methods, like 
spectral line addition, but this time we overcome the limitation of 
the weak-field approximation.
The first successful MZS tentatives consisted in the adding up 
of the circular polarisation of selected spectral lines. Paper~{\sc i}
(Semel 1989) recommended the de-blending of the spectral lines prior to
addition, but Semel \& Li (1996) showed that simple coherent addition 
of spectral lines gave satisfactory results. 
Some 200 spectral lines were added in that work to  yield significant 
detections.

This straightforward method of line addition was later improved by
Semel (1995). A list of {\em spectral lines of interest} was created.
These spectral lines were labelled by wavelength  $\lambda_i$,
equivalent width $w_i$ and effective Land{\'e} factor $g_i$. In the
next step, each spectral line was represented by a Dirac function
$w_i \cdot g_i \cdot \delta(\lambda - \lambda_i)$ and transformed, as
explained in Semel (1989), to the Doppler coordinate $X$, with
$dX = c \, d\lambda/\lambda$, and $c$ the velocity of light. The
convolution of the observed circular polarisation spectrum $V(X)$
and the line list results in a pseudo line $f(x)$ given by
\begin{equation}
f(x)= \Sigma_i\,(w_i\,g_i)\,V(x-X_i)/\Sigma_i\,(w_i\,g_i)
\end{equation}
As a rule, the convolution by means of a Fourier transform is found
to be quite efficient.
Donati et al. (1997) introduced the postulate that in a given
observed Stokes spectrum the shape of the Zeeman signature in
circular polarisation is identical for all spectral lines. Individual
Stokes $V$ profiles are assumed to correspond to the common basic
Zeeman signature, multiplied by the effective Land{\'e} factor and
by the central depth of the line in question. Based on this assumption,
a particular MZS (the basic Zeeman signature) is extracted by means of
a least squares method from the observed spectrum, In combination with
maximum entropy codes (Brown et al. 1991, Donati \& Brown 1997), LSD
based Stokes $I$ and $V$ profiles have been used for the production of
magnetic surface maps of quite a large number of stars with great success.

The methods and techniques used in solar polarimetry and magnetometry 
have subsequently been applied to the stellar problem. The solar case
is always the easier one, because one can often isolate just 
{\it one point} at a time and get strong enough signals for each 
wavelength pixel observed. Here the Doppler effect due to the solar 
rotation does NOT present any problem. On the contrary in the case 
of a star, the signals from all the points over the visible disc are 
mixed up. Also, the star being faint, the signals in its spectrum are 
much weaker and we need to collect the contributions over a
significant portion of the spectrum before we get reasonable signal
strengths. Moreover, the Doppler effect due to the stellar rotation 
cannot be neglected.

We therefore start our present discussion with solar conditions, i.e. a 
magnetic field observed in just one point and we shall take a few spectral lines formed in the presence 
of a magnetic field and calculate the Stokes vectors at spectral
resolutions common in solar physics, some 3\,millions. We then add  
up the Stokes parameters of all the spectral lines and subsequently
reduce the spectral resolution to stellar conditions, say 75000,
obtaining a kind of {\it pseudo-line}. We then examine the invariants 
that we have already found in solar magnetism, as the 
displacement of the centre of gravity that yields a good approximation 
to the longitudinal component of the magnetic field. Our purpose here
is to assess to what degree the new {\it pseudo-line} still contains
all the information on the magnetic field that was present in the 
individual spectral lines in our sample, well beyond the weak-field 
approximation. Such will be the first result of 
this work.

A similar procedure will be followed with the linear polarization signatures. 
Linear polarization is often neglected in stellar polarimetry seeking Zeeman 
signatures in the assumption that it will present amplitudes too small to 
be observed. The second result presented in this paper, the first of a series, 
will be the demonstration that, at 
appropriate spectral resolutions, the linear polarization due to the Zeeman 
effect produces sensible signatures useful for 
diagnostics of the stellar magnetic fields

In the following papers of this series we will apply recent solar inversion 
procedures
to the synthesised Stokes profiles, and we compare the results 
obtained with the original solar resolution with the {\it pseudo-line} 
smoothed to stellar resolution (paper II, Ramirez V/elez et al., 2008, 
submitted); then we shall replace the Milne-Eddington model atmosphere 
used in the 
radiative transfer computations by detailed line 
formation calculations in an empirical model atmosphere for a variety 
of magnetic field configurations. Application of principal component analysis 
(PCA) has been proven very effective 
in solar magnetometry and, analogously to the solar case, a database will be 
created and 
the PCA eigenvectors derived, applying the PCA-ZDI detection to just one 
magnetic point on the star. In addition we shall discuss how to apply our 
approach to some few points on the stellar disc well separated by 
significant Doppler effects (Paper III).
Finally in Paper IV we shall have to treat the more realistic case of 
continuous distributions of fields over the stellar surface. Here, 
the orthogonality of the eigenvectors is not conserved for adjacent 
stellar points (that is for small differences in the Doppler shifts).
The LSD technique may be proposed for such cases.

 \section{Invariants in Zeeman analysis.}

The effects of a magnetic field on the line profiles are quite specific
and essentially comprise polarisation and symmetry properties.
Both circular and linear polarisation appear in spectral lines:
linear polarisation is symmetrical in wavelength, circular 
polarisation on the other hand is antisymmetric. 
These manifestations are very particular to magnetic fields. For 
instance, only magnetic fields may create circular polarisation in 
spectral lines. The detection of such polarisation is unambiguous 
evidence for the presence of magnetic fields. It is our conviction that 
we can rely on some invariance properties in the spectropolarimetric 
data. Here we present a few specific arguments.

\subsection{ The centre of gravity method.}

This method applies only to circular polarisation; it is well studied 
in solar magnetism and described in a number of papers of which at
this point we mention only Rees et al. (1979)( and see references therein). In the limiting case
of optical thin layers in LTE, the centre of gravity shift of the 
line profiles observed in circular polarisation is proportional to
the longitudinal component of the magnetic field. This still holds 
true, even for line formation in optically thick layers, in the 
limiting case of weak magnetic fields. Usually, this simple method
is still a fair approximation in the more general case, but there 
are a few exceptions -- see Semel (1967) for theoretical 
demonstrations, experimental tests and some exceptions.

\subsection{The weak-field approximation.}

This approximation was first introduced by Sears, (1913) and was 
used for observations with solar magnetographs. It was the first 
assumption in the first paper on ZDI (Semel, 1989), henceforth 
paper\,I. Does this approximation also hold for stronger fields? 
In the following we shall present some relevant calculations 
concerning its application in the context of stellar magnetic 
field measurements.

Let us now analyse the weak-field approximation within the framework 
of the centre of gravity method. Consider the situation where the 
Stokes $V$ parameter is fully antisymmetric and $I$ is symmetric (in 
the absence of velocity gradients). When the weak field approximation 
holds, $V$ is nearly proportional to the first derivative of $I$, 
but this is no longer true for larger Zeeman shifts. However, if 
the spectrum is smoothed, the weak field situation is recovered 
to some extent and the weak-field approximation can be extrapolated 
to stronger fields. Indeed, smoothing neither affects the centre of 
gravity shifts nor the symmetry characteristics of $I$ or $V$, while 
the Zeeman shifts become small relative to the widths of the smoothed 
spectral lines. This will be shown explicitly below with a numerical 
demonstration. The weak-field approximation used in paper~I can thus 
still hold in stellar spectropolarimetry thanks to the low spectral 
resolution.

\subsection{Strong field - all Zeeman components separated.}

This is a simple case where magnetic measurements are possible with no 
reference to a model atmosphere since the polarisation states of all 
the components are completely determined by the orientation of the 
magnetic field. When the state of polarisation of any of the $\sigma$ 
components is known it is possible to determine the orientation of the 
field except for the ambiguity in the azimuth. The separation of the 
components determines the strength of the field.

\subsection{Inversion of full Stokes polarimetry.} 

Even the use of a simple M-E model can give reliable results provided 
that the parameters of the model are {\it not fixed} but determined
in the inversion procedure. Models more sophisticated than M-E could 
determine even gradients! All these approaches are improved by
multiline techniques.

\subsection{Our program}

We want to asses  and eventually develop the invariance in the magnetic 
fields measurements, so that the effects due to  poor predetermination  
of the atmosphere model are  almost negligible. For the time being,
our model will be limited to line formation in LTE, neglecting
vertical gradients.
 
We first present very simple 
examples \footnote{These steps are necessary for a better understanding} 
but at the end, using our approach with the PCA technique, ZDI will be 
pushed to its limits. To start with, we stick to the original meaning
of ZDI in its limited sense, i.e. magnetic fields occur only at 
``isolated''\footnote{to be explained later} points on the stellar
surface. Only much later will we tackle ZDI in the common and broader 
sense as the mapping of stellar surfaces.

\section{Multiline Zeeman Signatures in circular polarisation (MZSV) applied to the spectral lines of a single multiplet 
         of neutral iron.}

For this demonstration we selected the spectral lines of 
one specific multiplet of neutral iron (816) and we used a simple
Milne-Eddington (M-E) model, characterized by a linear with optical depth source function, keeping all other p
arameters constant with depth. The only other free parameters of the model will be the Doppler width, line damping, 
and opacity ratio of the line core respect to the continuum on the thermodynamical side of the line formation, 
a magnetic field vector and a line-of-sight velocity field. 

The lines of multiplet 816 of Fe\,{\sc i} have often been
used in solar magnetic field measurements. We keep in the present work a  high spectral 
resolution of 100\,m/s, i.e. a resolution of 3\,millions, and then 
study the effect of smoothing, bringing the resolution down to 75000,
which corresponds to sampling with a step of about 4\,km/sec. Without
loss of generality we employed a M-E model, using the code {\it Diagonal}
as described in L\'opez Ariste and Semel (1999) with only one layer. This code 
can be extended to a large number of layers and fit any LTE model. The 
atomic parameters are listed in Table\,1. For the M-E model we took 
$\beta = 10$ and $\eta$ equal to twice the relative strength of the 
components of the multiplet as given by Allen (1991). Doppler broadening 
is $21$\,mA as in sunspots, and damping is taken to be 0.01. 
\footnote{Saturation has always been an important issue in solar and 
stellar magnetic field measurements. Here it is included in the
procedure and we have to use the same procedure when we get to the 
inversion.}

\subsection{A justification of the M-E model, blending, and Li Jing's method}

Using the M-E model facilitates the analysis and does not require
heavy calculations. Calculating each line separately with a M-E model 
may appear subject to doubt, but the experience of solar magnetism has 
shown that when the parameters are not fixed but fitted to the 
observations, the M-E model is quite good. Selecting multiplet 816 of
Fe\,{\sc i} ensures the validity of {\bf LS} coupling and the easy
determination of Zeeman patterns, relative line strengths etc. 

Line blending seemed to be a serious matter in paper\,I (Semel,
1989). However, Semel and Li (1996) have shown that when numerous 
lines are ``added'' {\bf coherently}, the blending lines are added 
{\bf incoherently} and therefore only contribute to the noise
(which in turn is reduced when the number of spectral lines increases). 
In  Figs.\,1-5 we will show each profile in detail and one can see the 
limitation of the weak field approximation. Now, when we add the lines 
and then reduce the resolution we {\it recover} the weak-field
approximation (Fig. 6-10).

A comment on blending: each spectral line may appear several times, 
once in the coherent addition, and then each time when the line is 
``blending'' in the field of another nearby line. However, in the
latter case it appears in a non-coherent way and always in another 
position (in wavelength). When the line is blending, its contribution
is ``arbitrarily'' situated and has only a little effect.

\begin{table}
 \begin{tabular}{ccccccr}        \hline
&Line  & $\lambda$  & upper &   lower  & relative  & $\eta$\\
& No.  & \AA        & level &   level  & strength &       \\
 \hline
& 1  & 6400.010     & $^{5}D_4$   &    $^{5}P_3$   &    27     & 54   \\
& 2  & 6411.658     & $^{5}D_3$   &    $^{5}P_2$   &    14     & 28   \\
& 3  & 6408.031     & $^{5}D_2$   &    $^{5}P_1$   &    5.25   & 10.5 \\
& 4  & 6246.334     & $^{5}D_3$   &    $^{5}P_3$   &    7      & 14   \\
& 5  & 6301.515     & $^{5}D_3$   &    $^{5}P_2$   &    8.75   & 17.5 \\
& 6  & 6336.835     & $^{5}D_1$   &    $^{5}P_1$   &    6.75   & 13.5 \\
& 7  & 6141.734     & $^{5}D_2$   &    $^{5}P_3$   &    1.     & 2    \\
& 8  & 6232.661     & $^{5}D_1$   &    $^{5}P_2$   &    2.25   & 4.5  \\
& 9  & 6302.507     & $^{5}D_0$   &    $^{5}P_1$   &    3.     & 6    \\
\hline
\end{tabular}
\caption{The list of spectral lines of multiplet 816 of neutral iron.
The relative strengths correspond to LS coupling as given by Allen, 1991,
p. 62. The ratio $\eta$ of line to continuum absorption is obtained 
from the relative strengths multiplied by a factor of 2; the latter
is chosen arbitrarily.}
\end{table}

\begin{figure}[!htpb]
\resizebox{9cm}{!}{\includegraphics{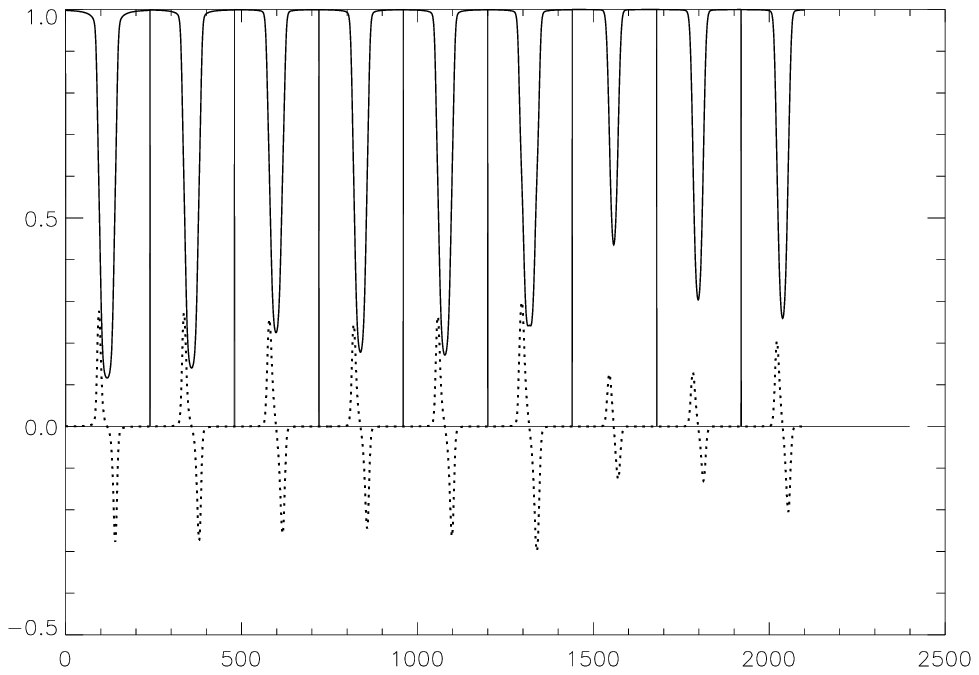}}
\caption{ The 9 lines, (line 1 to 9 from left to right) 
of multiplet 816 of neutral iron -- see Table\,1 -- put together with 
a spectral resolution of 3\,millions and sampling of 100\,m/sec,
corresponding to steps of about $2.1$\,m\AA in $\lambda$.  
Magnetic field with $B = 500$\,G and inclination $50^{\circ}$. The 
upper curves (continuous) show $I$ and the lower curves (dashed) $V$.}
\end{figure}

\begin{figure}
\resizebox{9cm}{!}{\includegraphics{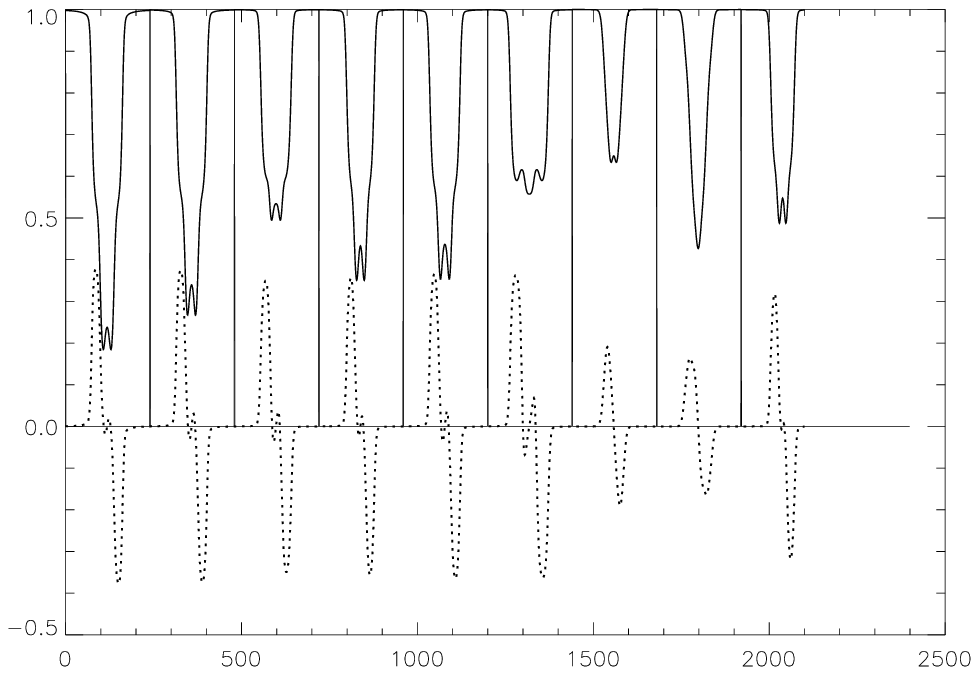}}
\caption{ The same as Fig.\,1 but with a field of 1500\,G.}
\end{figure}
\begin{figure}
\resizebox{9cm}{!}{\includegraphics{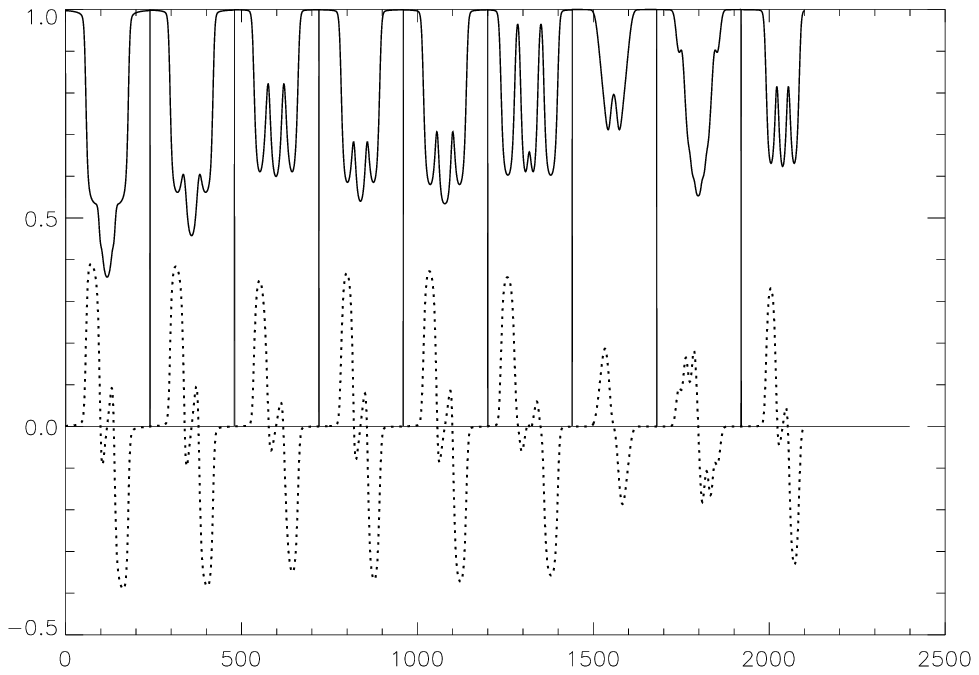}}
\caption{ The same as Fig.\,1 but with a field of 2500\,G.}
\end{figure}

\begin{figure}[!htpb]
\resizebox{9cm}{!}{\includegraphics{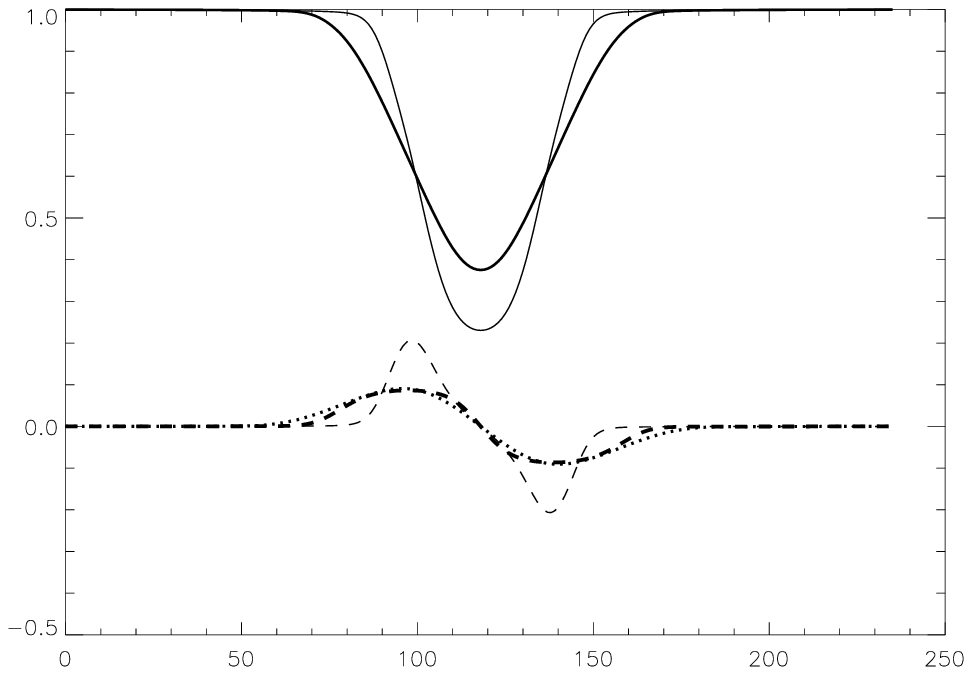}}
\caption{  The case of 500\,Gauss. 
The pseudo-line resulting from the addition of the
nine lines (thin continuous line), then smoothed by factor 40 to reduce 
the spectral resolution to 75000 (thick continuous). Bottom: The circular 
polarisation added (dashed) and smoothed (thick dashed). The derivative
of the smoothed intensity (points).}
\end{figure}

\begin{figure}[!htpb]
\resizebox{9cm}{!}{\includegraphics{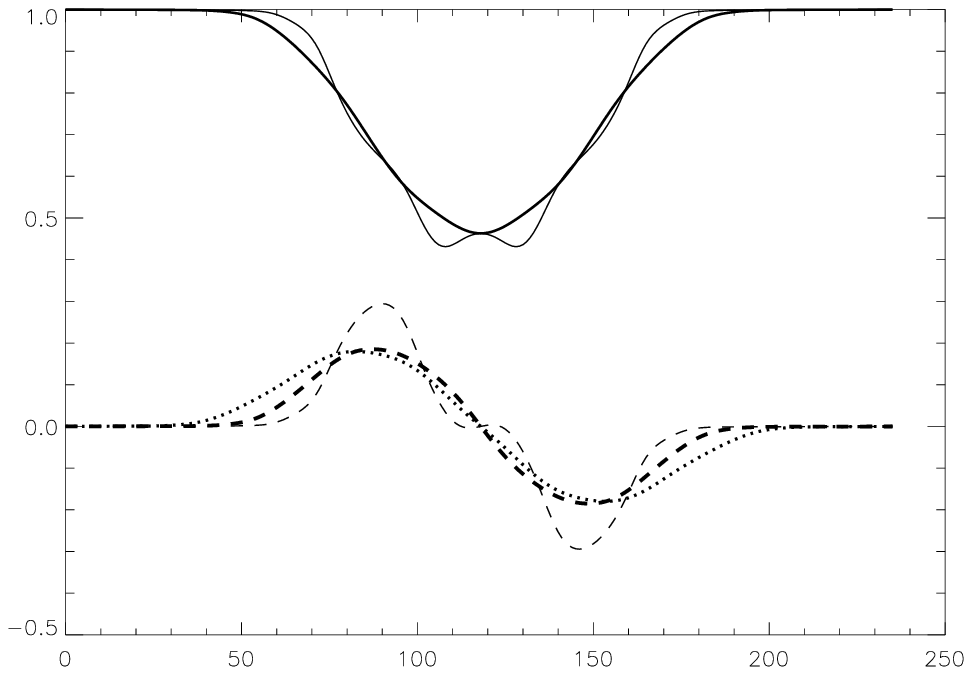}}
\caption{  The same as Fig.\,4, but for 1500\,Gauss.}
\end{figure}
\begin{figure}[!htpb]
\resizebox{9cm}{!}{\includegraphics{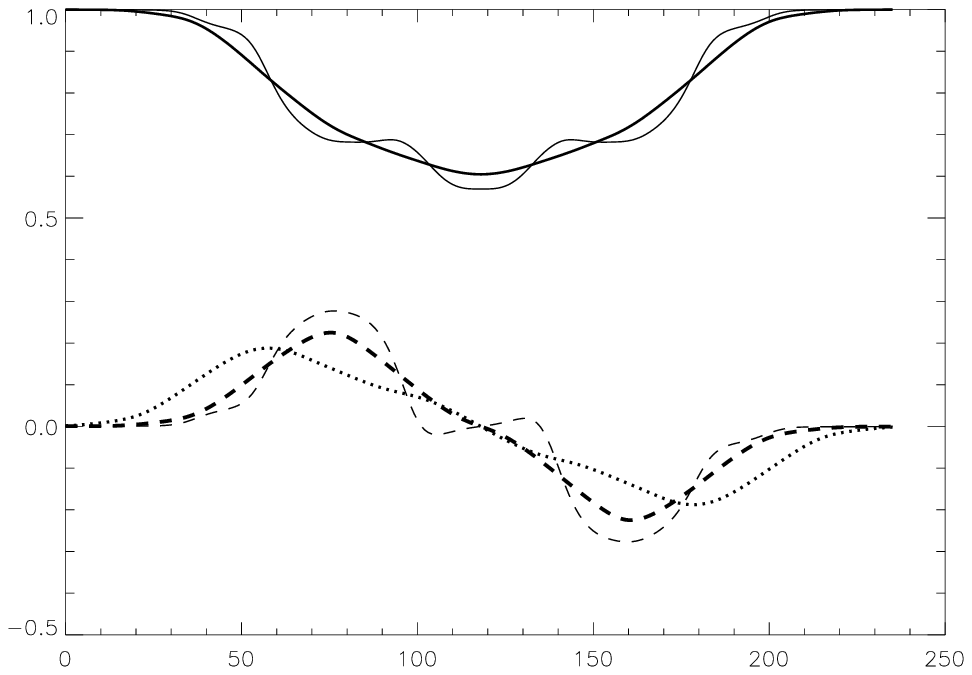}}
\caption{\textit{Top:} The same as Fig.\,4, but for 2500\,Gauss.}
\end{figure}

\subsection{Discussion of the figures}

 
With very high spectral resolution, (see Figs.\,1-3), the circular 
polarisation changes shape and details clearly depend on the field 
strength. For instance, the anomalous dispersions appear for fields 
stronger than 1\,kG as reflected in {\it the reversal} of Stokes $V$ 
at the line centres. This has no counterpart in the weak-field approximation.  

While we know from the solar physics that the centre of gravity method
really indicates the longitudinal component of the magnetic field, the 
application of the weak field approximation seems of doubtful validity
when we examine the individual line profiles in Figs. 1-3. However, 
the centre of gravity is a linear operator and commutes with the
algebra of line addition and with smoothing. 
So we may proceed to {\it operate} on the pseudo-lines ``observed'' in 
circular polarisation as shown in Figs. 4-6. In these 
figures, the curves for the Stokes profiles become smoother and one can
imagine that we eventually recover the essential features of the 
weak field approximation! We anticipate that with our approach we may
recover, in a stellar object, at least any astrophysical result that 
may be found through the weak-field approximation. More sophisticated 
methods of inversion will be described in the following papers.

\begin{figure}
\resizebox{9cm}{!}{\includegraphics{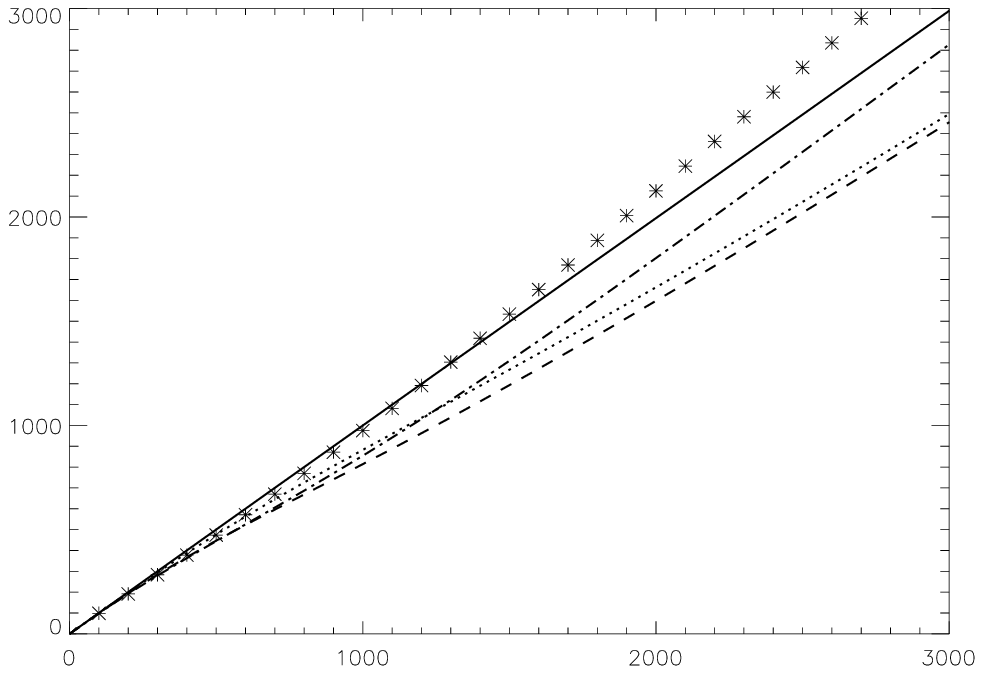}}
\caption{\textit{Top:}  A test of the centre of gravity method. The C.G. 
shifts are calculated over a database of 30 values of magnetic fields in 
steps of 100\,Gauss from 0 to 3\,kG. The shifts are divided by the cosine
of the inclination angle and by the adopted ``g~factor'' to compare with 
the amplitude of the magnetic field (abscissa). The curves are shown 
for the following values of ,
{\bf $\psi$}
,the inclination angles in degrees:
$0$ (solid line), 20 (points),  40 (dashed), 60 (dashed - points)
and 80 (stars). The ``g~factor'' for the pseudo-line was chosen to be
1.59, For $\psi = 0$ the C.G. shifts correspond  exactly to the line
of sight component of the magnetic field. 
}
\end{figure}
\subsection{Crtical discussion of the C.G. method, unresolvzd magnetic 
fields and Fig. 7}
 Even the best solar telescope can not resolve the magnetic element 
in a solar faculae. One way to tackle the problem was to turn to a 
kind of the  {\it filling factor algebra.} An account is given in 
(Semel,1981).  With the help of the C.G. method one could calculate the {\it relative magnetic  curve of growth.}

 {\bf However, it is beyond the scop of this paper. In short, Fig.7 is 
symbolic and not a solution of the problem. The C.G. may be a good 
tool to start,  but the real work must be followed by an inversion code.
(Landstreet,2008) warns us that the errors in magnetic field determination  
by use of Fig. 7 reach  several hundreds gauss.
 While the C.G. method was useful at early time, for the determination 
 of solar magnetic fields, it is not recommended today.
 In Fig. 7, we wanted to show that the pseudoline contains a Zeeman 
signatures of the LOS component within 95 percents of certainty.  
The appropiate  
inversion will be the object of the next paper.}
 For further discussion of the C.G. method applied to stars see 
(Stift M.,1986)  and (Leone F. \& Catanzaro 2004)

\subsection{Conclusions for the first part: Zeeman circular polarisation}

The pseudo-lines are not simple spectral lines in the sense that they have 
no wavelength and are not characterised by specific atomic parameters 
such as the excitation potential. Moreover, in the more general case they 
do not even correspond to a particular chemical element, and no specific 
spectroscopic term is attached to them. Still they contain a lot of 
information on the astronomical objects we are interested in. What is
important: the Zeeman signatures do not disappear in the pseudo-lines! 
Giving up the weak field approximation, we may look for more sophisticated 
methods of determining magnetic fields. Applying for example PCA
procedures, we can obtain more parameters than with methods limited by
the weak field approximation.

\section{Spectral resolution and the observation of linear Zeeman
  polarisation in solar type stars}

Carroll et al. (2007) state that as a rule of the thumb, the Zeeman linear
polarisation is at least one order of magnitude lower than the circular one.
Now, if our line of sight (henceforth LOS) is not a preferred direction 
for the stellar magnetic field and therefore its three components  
(the longitudinal one and the two perpendicular to the LOS) have all equal 
probabilities, than on the average, the transverse component is $\sqrt2$
times stronger than the LOS component. The average of the angle between 
the magnetic vector and the LOS should thus be 
$\approx arc cos(1/\sqrt 3) \approx 54^{\circ}$.
From solar physics we know that kG fields are not only found in sunspots. 
But can we assume that Kilogauss fields are common also in solar type 
stars? If yes, why does the linear polarisation escape observation? In 
the following, we try to answer this question and we also suggest a remedy.

 \subsection{ The most probable angle between the LOS 
and the local magnetic vector.}

Let us call this angle ${\psi}$ and take it between $0^{\circ}$ 
and $90^{\circ}$, say positive LOS component. The average $\overline{\psi}$ 
is 1\,radian if all directions have the same probability; the median angle 
is  $\psi_{Median} = 60^{\circ}$. In other words, there are as many orientations 
between $0^{\circ}$ and $60^{\circ}$ as between $60^{\circ}$ and $90^{\circ}$. 
For the case of negative LOS longitudinal fields,
$90^{\circ}<{\psi}< 180^{\circ}$, the results for  $\overline{\psi}$ and for  
$\psi_{Median}$ are deduced similarly. In conclusion, $\psi$ is likely to 
be nearly $60^{\circ}$. In the following we define the Stokes parameter
$Q$ as perpendicular to the magnetic field. We will show now that $Q$
is likely to be significant. In the next figures we see that for
1\,kG and $\psi = 60^{\circ}$, MZSQ $\approx 0.7$ MZSV for high spectral 
resolutions. Now $Q$ changes with $\lambda$ twice faster than $V$, so that  
for current high spectral resolutions, say 60000, $Q$ shrinks much more than
 $V$. 
Moreover, $Q$ changes sign when the azimuth changes by $90^{\circ}$, while $V$ 
changes sign when $\psi$ changes by $180^{\circ}$. Both effects may be 
reduced by increasing spectral resolutions, say $120000$. 
As is seen in Fig.\,8,
$Q$ is preserved considerably and spatial resolution of the stellar surface is 
improved as well; this is shown in Figs.\,12-17.
 
 \begin{figure}[!htpb]
\resizebox{9cm}{!}{\includegraphics{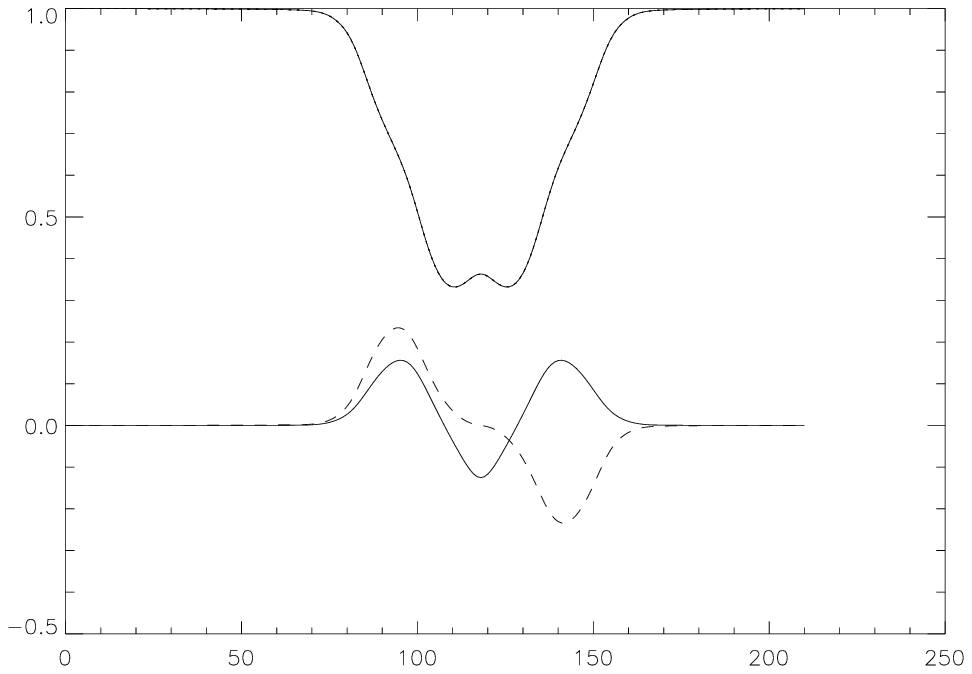}}
\caption {The Stokes profiles, $I,Q,V$ of the pseudo-line
for a magnetic field $H = 1000$\,Gauss, $\psi = 60^{\circ}$ with the 
original resolution of 3\,millions (thin). $I$ (points) and $Q$
(continuous), $V$ (thin).} 
\end{figure}
\begin{figure}[!htpb]
\resizebox{9cm}{!}{\includegraphics{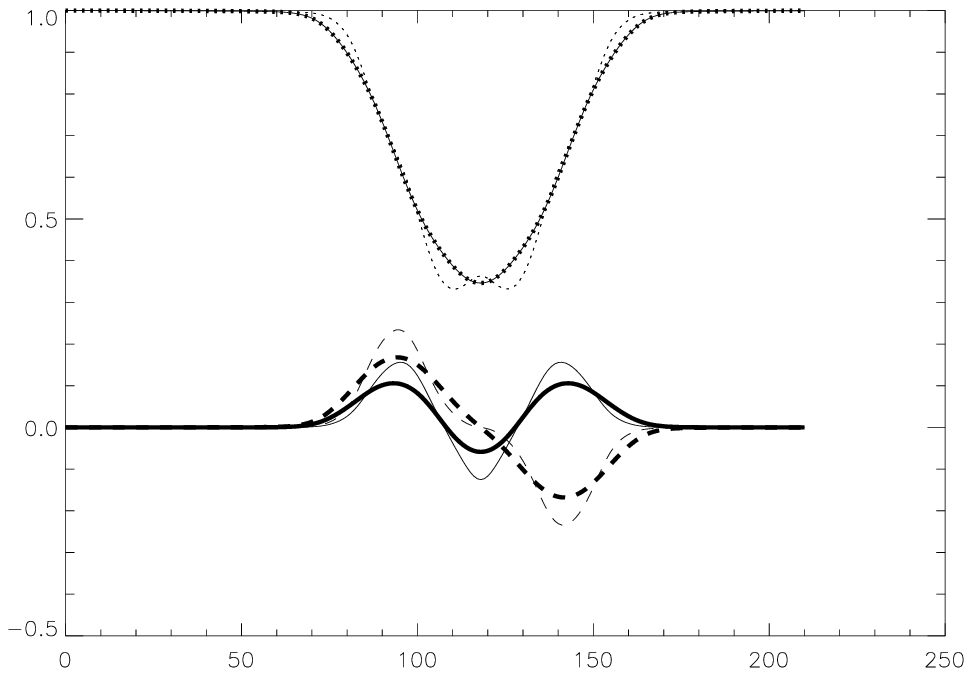}}
\caption{ I,Q,V. The same field as before, original resolution 3\,millions
(thin).  Resolution 120000 (thick). Note that at this field strength
(1000\,Gauss) $Q$ is still detectable with this resolution.}
\end{figure}
\begin{figure}[!htpb]
\resizebox{9cm}{!}{\includegraphics{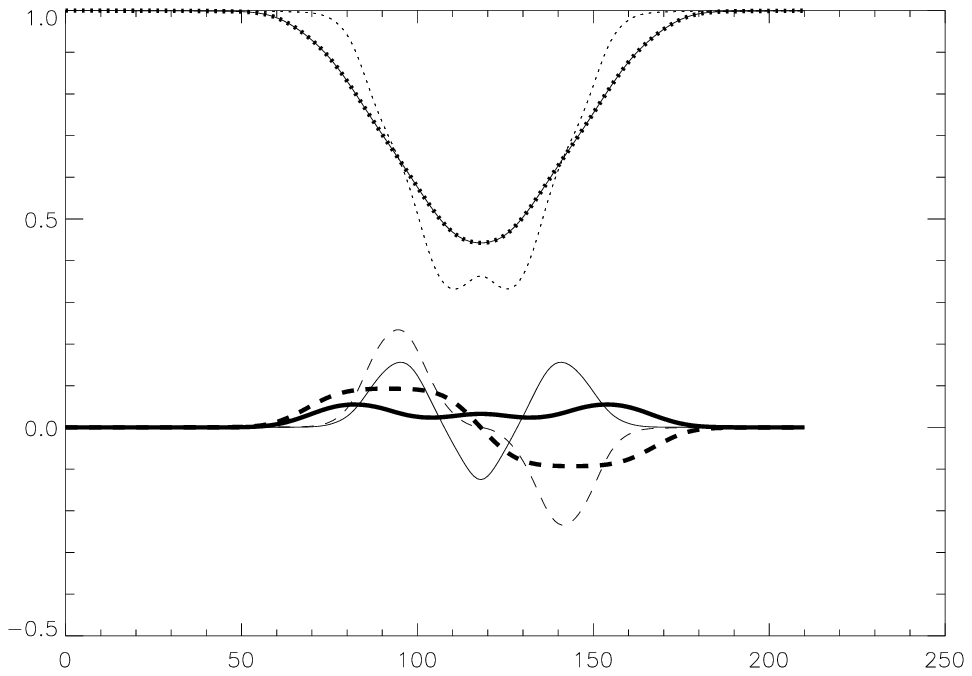}}
\caption{ I,Q,V. The same field as before, original resolution 
3\,millions (thin). Resolution 60000 (thick); Note that $Q$ is reduced 
considerably for 1000\,Gauss with the low resolution.}
\end{figure}
\begin{figure}[!htpb]
\resizebox{9cm}{!}{\includegraphics{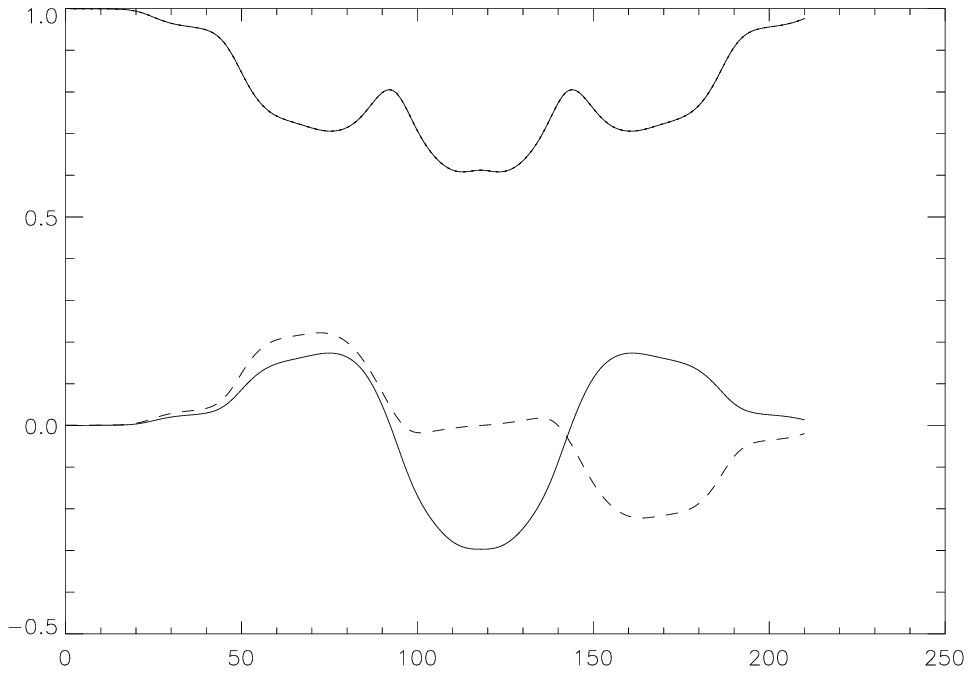}}
\caption{I,Q,V. $H = 3000$\,Gauss. Original resolution (Thin).  
}
\end{figure}
\begin{figure}[!htpb]
\resizebox{9cm}{!}{\includegraphics{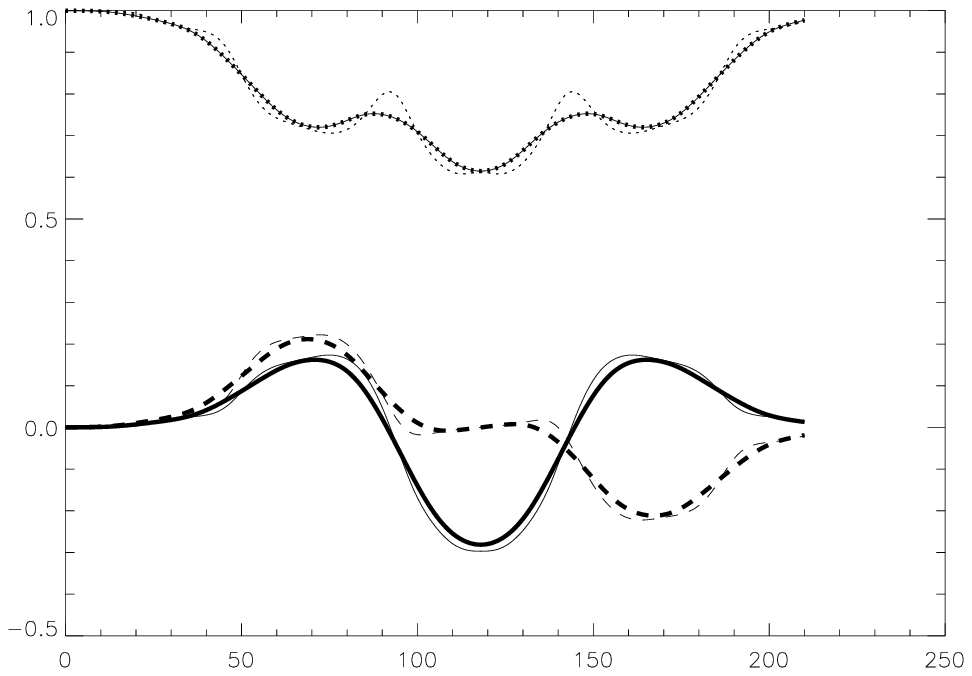}}
\caption{AS in the last figure but with resolution 120000 (thick)}
\end{figure}
\begin{figure}[!htpb]
\resizebox{9cm}{!}{\includegraphics{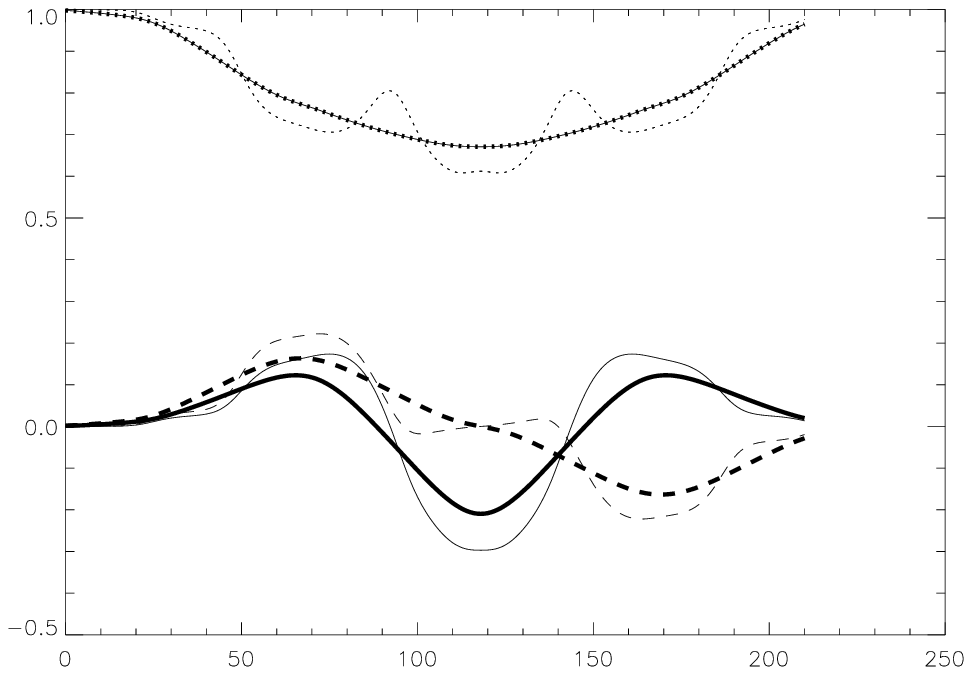}}
\caption{ AS in the last figure but with resolution 60000 
(thick) Note that for field as high as 3000\,Gauss, $Q$ is still 
detectable even with 60000 resolution} 
\end{figure}

\subsection{Conclusions for the second part: Zeeman linear polarisation}

 We have used the pseudo-line technique to show how to extract a Zeeman 
signature for linear polarisation in a way similar to what we did in
the first part for circular polarisation. The first conclusion concerns 
the spectral resolution required.
With a resolution superior to 100000, we can probably detect significant 
signals in linear polarisation in the spectra of solar type stars!
Such a resolution has been achieved in spectroscopic tests with UCLES at the
AAT with the 79\,G/mm grating (see, for instance, Lopez et al. 1999).
A single observation of linear polarisation in the spectrum 
of a solar type star has been  performed in 2004 (see Semel et al, 2006). The
linear signal observed was indeed four time less than the circular one, 
but still significant. There is definitely an interest to proceed in 
this direction.  

\acknowledgements
Dr. David Rees has introduced PCA methods to the field of solar magnetometry 
which was the starting point for this method in general magnetometry 
including solar type stars.
Here,  one of us (M. Semel) wishes to express his gratitude
to Dr. David Rees for lectures on PCA given during his visit to Meudon 
Observatory in the year 2002. \\
MJS acknowledges support by the {\sf\em Austrian Science Fund (FWF)}, 
project P16003-N05 ``Radiation driven diffusion in magnetic stellar
atmospheres'' and through a Visiting Professorship at the Observatoire
de Paris-Meudon and Universit{\'e} Paris 7 (LUTH).
Our thanks go to Prof. S. Cuperman for reading the paper and for very 
precious comments and corrections.

\section {References} 
Allen, C.W.,1991, Astrophysical Quantities, Publisher ATHLONE PRESS LTD.\\
Carrol T.A.,Kopf,M., Ilyn L. and Strassmmeier K.G., 2007, Astron. Nachr./AN 999,
{\bf 88},789. \\
Brown, S. F.; Donati, J.-F.; Rees, D. E. and Semel, M.,1991 Astron. Astrophys. {\bf250},463\\
Donati, J.-F.; Brown, S. F. 1997,Astron. Astrophys. {\bf 326},1135\\
Donati, J.F. et al. 1997, MNRAS,{\bf 291},658 \\
Landstreet,2008 (private communication)\\
Leone F. \& Catanzaro 2004, Astron. Astrophys.,425,271.\\
L\'opez A. and Semel M., 1999, Astron. Astrophys. Suppl. Ser. {\bf 139}, 417.\\
L\'opez,A. and Semel M.,1999, WEB page :  www.aao.gov.au   /local/www/UCLES/cookbook/report2.html.\\
Rees, D. E. and Semel, M. D., Astron. Astrophys., 1979,{\bf 74},1. \\
Rees, D. E., L\'opez Ariste A.,Thatcher J., and Semel M., 2000,Astron. Astrophys., {\bf 355},759. \\
Sears F. H. 1913, Ap. J. {\bf 38},99.\\
Semel,M.,1967,Astron.Astrophys.,{\bf 30},257. \\
Semel, M., Astron. Astrophys.,1980, {\bf 91}, 369.\\
Semel, M., Astron. Astrophys.,1987, {\bf 178}, 257.\\
Semel  M.,1989,Astron. Astrophys.,{\bf 225},456. \\
Semel M.and al. Third Afcop meeting, Toulouse 1995, editor  J.-F.  Donati. \\
Semel M. and Li J.,1996,Solar Phys.,{\bf 164},417.\\
Semel M.,Rees D.E.,Ramirez V\'elez,J. C.,Stift, M.J.and Leone F.,2006,(spw4),
ASP Conference Series,  vol. 358, eds. R. Cassini and Lites B. W. \\
Stift M.J.1986, MNRS 221,499\\
Stift,~M. J. 2000, COSSAM: Codice per la Sintesi Spettrale
           nelle Atmosfere Magnetiche, A Peculiar Newsletter, 33, 27\\

\end{document}